\documentclass[multphys,vecphys]{svmult}

\usepackage{makeidx}         
\usepackage{graphicx}        
\usepackage{multicol}        
\usepackage[bottom]{footmisc}

\makeindex             

\begin{document}

\title*{Multiplicity of chemically peculiar stars}
\titlerunning{Multiplicity of CP stars}
\author{Swetlana Hubrig\inst{1}\and
Nancy Ageorges\inst{1}\and
Markus Sch\"oller\inst{1}}
\institute{European Southern Observatory (ESO), Casilla 19001, Santiago, Chile}
%
%

\maketitle

\begin{abstract}
Recently, with the goal to study multiplicity of chemically peculiar stars,
we carried out a survey of 40 stars using diffraction
limited near infrared (IR) imaging with NAOS-CONICA (NACO) at the VLT. Here, we announce
the detection of 27 near IR companion candidates around
25 late B-type chemically peculiar stars exhibiting strong overabundances 
of the chemical elements Hg and Mn in
their atmospheres. A key point for the understanding of the abundance
patterns in these stars may be connected with binarity and multiplicity.
It is intriguing that more than half of the sample of HgMn stars studied
previously by speckle interferometry and recently using the adaptive optics
system NACO belong to multiple systems. 

\end{abstract}

\section{Introduction}
\label{hubrig:sec1}

Chemically peculiar (CP) stars are main sequence A and B type stars in the spectra 
of which lines of some elements are abnormally strong or weak. 
The class of CP stars is represented by roughly three subclasses 
including magnetic Ap and Bp stars, metallic-line Am stars and HgMn stars which
are late B-type stars
showing extreme overabundances of Hg (up to 6\,dex) and/or Mn (up to 3\,dex).
Among the magnetic Ap stars the rate of binaries is 43\% \cite{hubrig:car2002}.
The main result of this most recent study of multiplicity of magnetic stars 
is that, statistically, the orbital parameters of Ap stars do not differ from those of 
normal stars, except for an almost complete lack of orbital periods shorter than 
3\,days. The studies of the evolutionary state of magnetic Ap stars in binaries indicate 
that all of them are rather old main sequence stars and are well evolved from the 
zero-age main sequence (e.g.\ Wade et~al.\ (1996) \cite{hubrig:w1996}), fully in agreement with
results of the study of single magnetic Ap stars by Hubrig et~al.\ (2000) \cite{hubrig:h2000}.

The number of double-lined spectroscopic binary (SB2) systems among magnetic Ap stars
is abnormally low (only 3 SB2 systems are known to date) and no eclipsing binary 
comprising a magnetic Ap star has ever been identified. 
The rate of binaries is much smaller among magnetic Bp stars, $\sim$20\%,
and only two
double-lined eclipsing binaries have recently been discovered:
HD\,123335 was discussed by Hensbergen et~al.\ (2004) \cite{hubrig:he2004}
and AO\,Vel by Gonz\'alez et~al.\ (2005) \cite{hubrig:go2005}.
A total of six magnetic Bp/Ap stars are known to be multiple and are listed in the Multiple 
Star Catalogue of Tokovinin (1997) \cite{hubrig:to1997}.

The metallic-line Am stars show an overabundance of heavy elements and an underabundance of
Ca and Sc. A very high fraction of these stars, at least 90\%, are SB systems with orbital
periods between 2.5 and 100~days.

HgMn stars are rather young objects and many of them are found in young associations
like Sco-Cen, Orion OB1 or Auriga OB1. HgMn stars do not have 
strong large-scale organized magnetic fields and
exhibit marked abundance anomalies of numerous elements. In contrast to Bp and Ap 
stars with large-scale organized magnetic fields, 
they generally do not show overabundances of rare earths, but exhibit strong
overabundances of heavy elements like W, R, Os, Ir, Pt, Au, Hg, Tl, Pb, Bi.
Another important distinctive
feature of these stars is their slow rotation 
($\langle v\,\sin i\rangle \approx$ 29\,km\,s$^{-1}$, Abt et~al.\ (1972) \cite{hubrig:abt1972}).
The number of HgMn stars decreases sharply with increasing rotational
velocity\,\cite{hubrig:wo1974}. Evidence that stellar rotation
does affect 
abundance anomalies in HgMn stars is provided by the rather sharp cutoff in 
such anomalies at a 
projected rotational velocity of 70--80~km$\,$s$^{-1}$\,\cite{hubrig:hm1996}.

The mechanisms responsible for
the development of the chemical anomalies of HgMn stars are not
fully understood yet. 
A key point for the understanding of the abundance patterns may be connected
with binarity and multiplicity. More than 2/3 of the HgMn stars are known
to belong to spectroscopic binaries\,\cite{hubrig:hm1995}.
Quite a number of HgMn stars belong to triple
or even quadruple systems\,\cite{hubrig:co1992,hubrig:iso1991}. 
Out of 30 SB
HgMn stars observed in speckle interferometry, 15 appear to have more than two
components.
Indirect evidence for the presence of a third component was
found in four other HgMn SBs (HD\,11905, HD\,34364, HD\,78316 and HD\,141556) on the
basis of spectroscopic and photometric arguments.
Further evidence that other HgMn stars frequently are members of
multiple systems is inferred from the results of the ROSAT all-sky
survey. X-ray emission was detected through this survey in 12
HgMn stars (7 SB1s, 3 SB2s and 2 for which no radial velocity data are
available). Previous X-ray observations with the Einstein Observatory and
theoretical estimates had suggested that stars in the spectral range
B2--A7 are devoid of any significant X-ray emission.  In most cases
when emission had been detected in such stars, it was found to
originate from a cool companion.
This suggests that the X-ray
emission found in HgMn SBs does not originate from the HgMn primary.
From observations investigating late-B X-ray
sources using the ESO 3.6-m with ADONIS\,\cite{hubrig:hu2001}
we found faint companions for 4 HgMn stars that were part of the X-ray selected
late-B stars observed, strengthening this interpretation.  

In the catalogue of multiple stars by Tokovinin (1997) \cite{hubrig:to1997},
which compiles data on 612 stellar systems of different spectral types, we
found four additional multiple systems containing HgMn stars.
It is especially intriguing that if
the relative frequency of HgMn stars in multiple systems
is studied, every third system
with a primary in the spectral range between B6 and B9 involves an HgMn star.
These observational results clearly show that the study of
multiple systems with an HgMn component is of prime interest
for star formation.
In the following we report our results of the recent study of multiplicity
of this amazing class of objects using NACO K-band imaging.
  
\section{Observations}
Observations of 40 HgMn stars have been carried out with NACO in service mode 
from October 2004
to March 2005. We used the S13 camera, to be able to discover practically all 
components down to K=14 with a signal-to-noise ratio of the order of 12. 

Here, we announce the detection of 27 near IR companion
candidates in eight binaries, 16 triple and one
quadruple system. The detected companion candidates
have K magnitudes 
between 5${\hbox{$.\!\!^{\rm m}$}}$5 and 13${\hbox{$.\!\!^{\rm m}$}}$5 and angular 
separations ranging from 0${\hbox{$.\!\!^{\prime\prime}$}}$1
to 7${\hbox{$.\!\!^{\prime\prime}$}}$3 (7-1500\,AU).

\begin{figure}
\begin{center}
\begin{tabular}{|c|c|c|c|}
\hline
\resizebox{0.22\textwidth}{!}{\includegraphics{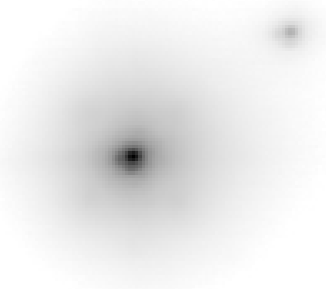}} &
\resizebox{0.22\textwidth}{!}{\includegraphics{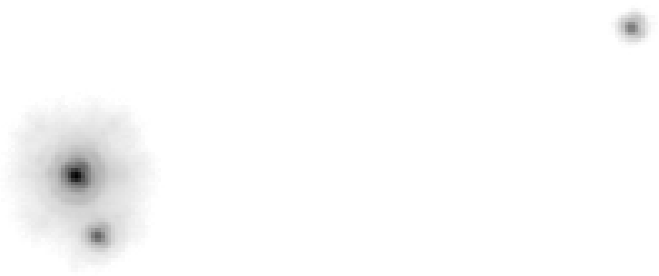}} &
\resizebox{0.22\textwidth}{!}{\includegraphics{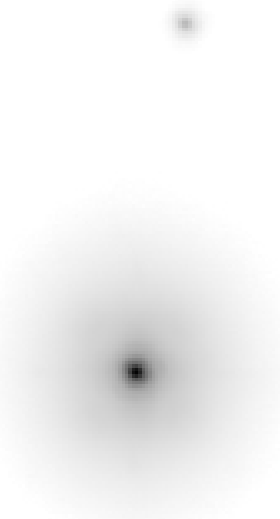}} &
\resizebox{0.22\textwidth}{!}{\includegraphics{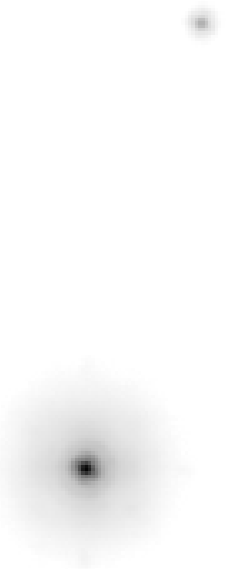}} \\
\hline
\resizebox{0.22\textwidth}{!}{\includegraphics{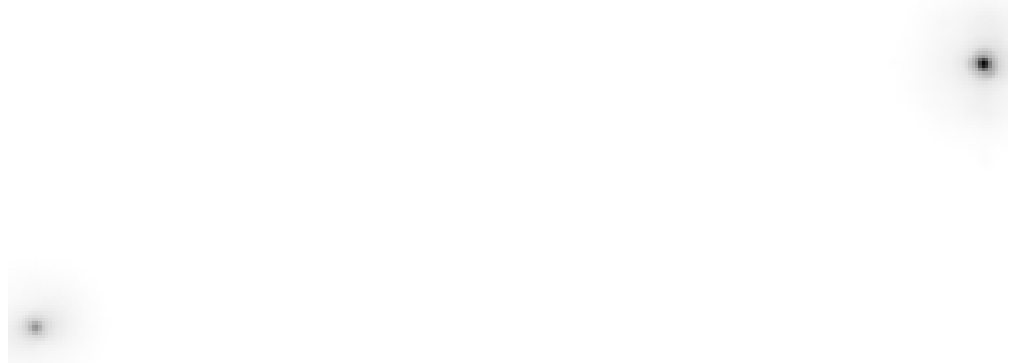}} &
\resizebox{0.22\textwidth}{!}{\includegraphics{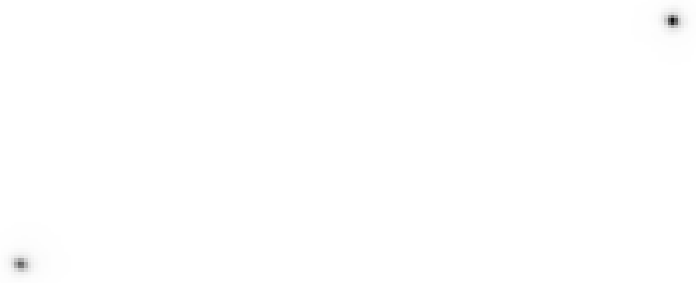}} &
\resizebox{0.22\textwidth}{!}{\includegraphics{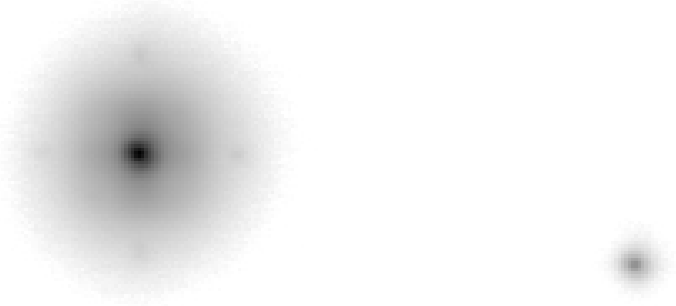}} &
\resizebox{0.22\textwidth}{!}{\includegraphics{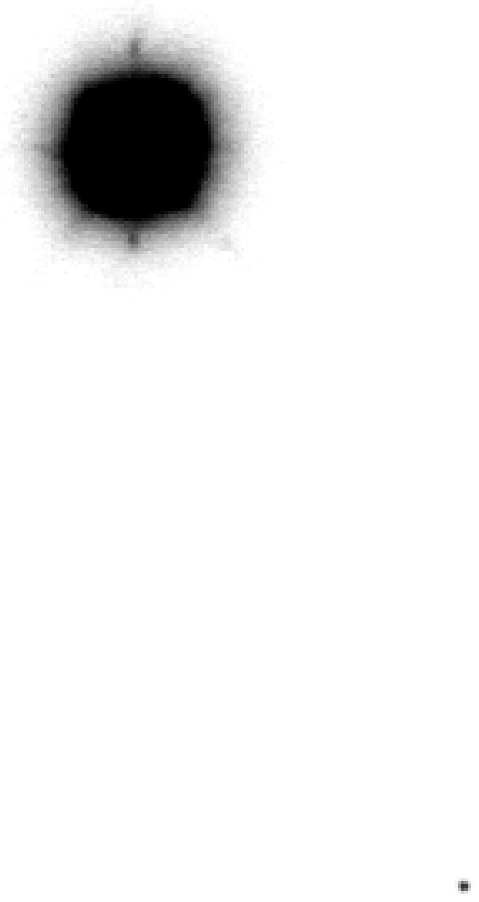}} \\
\hline
\end{tabular}
\caption{
NACO images of the wide systems in our sample.
Upper row from left to right:
HD\,32964, HD\,34880, HD\,36881 and HD\,53929.
Lower row from left to right:
HD\,120709, HD\,129174, HD\,165493 and HD\,178065.
The field of view in each frame is
7.95$^{\prime\prime}$$\times$7.95$^{\prime\prime}$.
}
\label{hubrig:fig1}
\end{center}
\end{figure}

\begin{figure}
\begin{center}
\begin{tabular}{|c|c|c|c|c|}
\hline
\resizebox{0.18\textwidth}{!}{\includegraphics{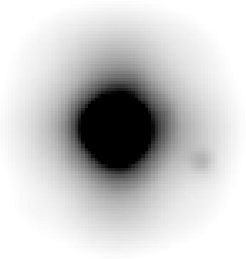}} &
\resizebox{0.18\textwidth}{!}{\includegraphics{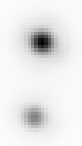}} &
\resizebox{0.18\textwidth}{!}{\includegraphics{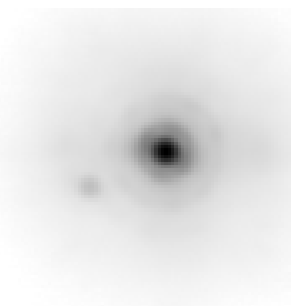}} &
\resizebox{0.18\textwidth}{!}{\includegraphics{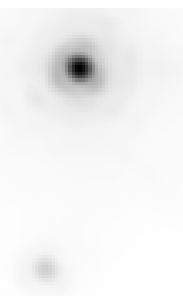}} &
\resizebox{0.18\textwidth}{!}{\includegraphics{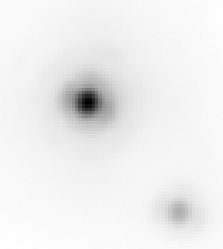}} \\
\hline
\resizebox{0.18\textwidth}{!}{\includegraphics{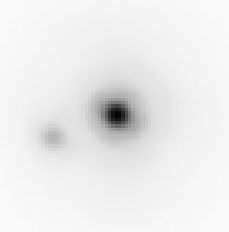}} &
\resizebox{0.18\textwidth}{!}{\includegraphics{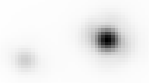}} &
\resizebox{0.18\textwidth}{!}{\includegraphics{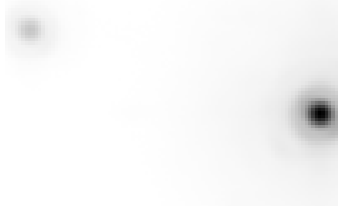}} &
\resizebox{0.18\textwidth}{!}{\includegraphics{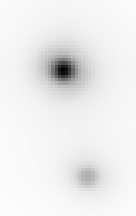}} &
\resizebox{0.18\textwidth}{!}{\includegraphics{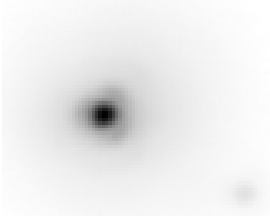}} \\
\hline
\end{tabular}
\caption{
NACO images of the intermediate systems in our sample.
Upper row from left to right: 
HD\,33904, HD\,35548, HD\,53244, HD\,59067 and HD\,73340.
Lower row from left to right: 
HD\,78316, HD\,101189, HD\,110073, HD\,158704 and HD\,221507.
The field of view in each frame is
1.33$^{\prime\prime}$$\times$1.33$^{\prime\prime}$.
}
\label{hubrig:fig2}
\end{center}
\end{figure}

\begin{figure}
\begin{center}
\begin{tabular}{|c|c|c|}
\hline
\resizebox{0.3\textwidth}{!}{\includegraphics{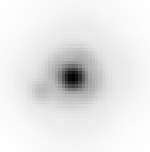}} &
\resizebox{0.3\textwidth}{!}{\includegraphics{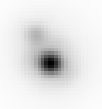}} &
\resizebox{0.3\textwidth}{!}{\includegraphics{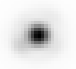}} \\
\hline
\resizebox{0.3\textwidth}{!}{\includegraphics{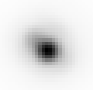}} &
\resizebox{0.3\textwidth}{!}{\includegraphics{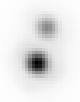}} &
\resizebox{0.3\textwidth}{!}{\includegraphics{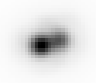}} \\
\hline
\end{tabular}
\caption{
NACO images of the close systems in our sample.
Upper row from left to right: 
HD\,21933, HD\,28217 and HD\,29589.
Lower row from left to right: 
HD\,31373, HD\,33647 and HD\,216494.
The field of view in each frame is
0.66$^{\prime\prime}$$\times$0.66$^{\prime\prime}$.
}
\label{hubrig:fig3}
\end{center}
\end{figure}

In Figs.\,\ref{hubrig:fig1}--\ref{hubrig:fig3} we show adaptive optics K-band images obtained 
with NACO.
The field of view displayed was selected according to the angular distances 
of the companions.
The intensity scale was adjusted to visualize the respective companions.

\begin{figure}
\begin{center}
\resizebox{0.6\textwidth}{!}{\includegraphics{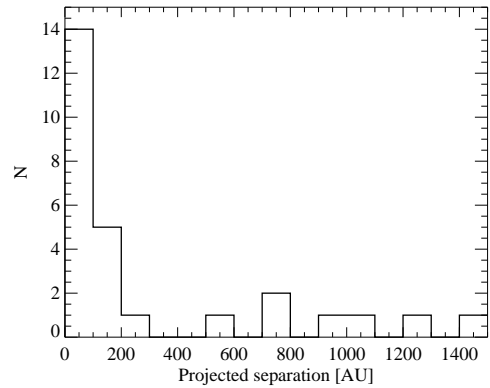}}
\caption{
Distribution of the projected separations for the studied systems with 
HgMn primaries.
}
\label{hubrig:fig4}
\end{center}
\end{figure}

One of our NACO targets, the
HgMn star HD\,75333, has already been observed in May 2001 with the adaptive optics
system at Keck\,II.
The Keck images of this system have been presented in Hubrig et~al.\ (2005) \cite{hubrig:hu2005}.
These observations revealed that this star has
two low-mass companions in a binary system
at a separation of 1${\hbox{$.\!\!^{\prime\prime}$}}$34.
This system is not displayed in Figs.\,\ref{hubrig:fig1}--\ref{hubrig:fig3}, since
the separation between the two low-mass companions in the binary system
is only 0${\hbox{$.\!\!^{\prime\prime}$}}$06. 
The diffraction
limit for NACO installed at the 8\,m telescope is lower than that for Keck\,II, 
hence these companions do not appear resolved in our NACO images. 
In Fig.\,\ref{hubrig:fig4} we show the distribution of the projected separations for the studied
multiple systems with HgMn primaries. For most of the systems the separations are 
smaller than 100\,AU.
 
If all detected IR objects around the HgMn stars are true companions, the resulting
multiplicity rate is 68\%. In Table\,\ref{hubrig:tab1} we present the list of the observed 
HgMn stars. Their visual magnitudes and spectral types were retrieved from the SIMBAD data 
base. In the last column we give some remarks about their multiplicity.

\begin{table}
\begin{center}
\caption{
List of HgMn stars observed with NACO.
}
\begin{tabular}{rccl|rccl}
\hline
\multicolumn{1}{c}{HD} &
\multicolumn{1}{c}{V} &
\multicolumn{1}{c}{Sp.\ Type} &
\multicolumn{1}{c}{Remarks} &
\multicolumn{1}{c}{HD} &
\multicolumn{1}{c}{V} &
\multicolumn{1}{c}{Sp.\ Type} &
\multicolumn{1}{c}{Remarks} \\
\hline
1909 & 6.6 & B9IV & SB2 &
70235 & 6.4 &B8Ib/II &  \\
7374 & 6.0 & B8III& SB1&
71066 & 5.6& A0IV& vis.\ binary \\
19400 & 5.5 &B8III & vis.\ binary &
71833 & 6.7 & B8II & vis.\ binary \\
21933 & 5.8 & B9IV & IR comp. &
72208 & 6.8 & B9 & SB2 \\
27295 &  5.5 & B9IV & SB1 &
73340 & 5.8 & B8 & IR comp. \\
28217 & 5.9 & B8IV & vis.\ binary + IR comp.&
75333 & 5.3 & B9 & two IR comp. \\
29589 & 5.4 & B8IV & SB1 + IR comp.&
78316 & 5.2 & B8III & SB2 + IR comp. \\
31373 & 5.8  & B9V & IR comp.&
101189 & 5.1 & B9IV & IR comp.\\
32964 & 5.1 & B9V & SB2 + IR comp.&
110073 & 4.6 & B8II/III & SB1 + IR comp. \\
33904 & 3.3 & B9IV & IR comp. &
120709 & 4.6 & B5III & IR comp. \\
33647 & 6.7 & B9V & SB2 + IR comp. &
124740 & 7.9 & A & SB2 \\
34880 & 6.4 & B8III & SB1 + two IR comp. &
129174 & 4.9 & B9 & SB1 + IR comp. \\
35548 & 6.6 & B9 & SB2 + IR comp. &
141556 & 4.0 & B9IV & SB2\\
36881 & 5.6 & B9III & SB1 + IR comp. &
144661 & 6.3 & B8IV/V & SB1 \\
49606 & 5.9 & B7III & SB1 &
144844 & 5.9 & B9V & SB2 \\
53244 & 4.1 & B8II & SB1 + IR comp. &
158704 & 6.1 & B9II/III & SB2 + IR comp. \\
53929 & 6.1 & B9.5III & SB1 + IR comp. &
165493 & 6.2 & B7.5II & SB1 + IR comp. \\
59067 & 5.9 & B8 & IR comp. &
178065 & 6.6 &B9III  & SB1 + IR comp. \\
63975 & 5.1 &B8II  & SB1 &
216494 & 5.8 & B8IV/V & SB2 + IR comp. \\
65950 & 6.9 & B8III &  &
221507 & 4.4 & B9.5IV & IR comp. \\
\hline
\end{tabular}
\label{hubrig:tab1}
\end{center}
\end{table}

\section{Discussion}
The results of our study clearly confirm that HgMn stars are frequently 
found in multiple systems.
It is especially intriguing that out of the 40 HgMn stars
in the sample studied only two stars, HD\,65950 and HD\,70235, are not known
to belong to a binary or multiple system. 
However, companionship can not be established based on K photometry alone, and
confirming the nature with a near infrared spectrograph is essential for 
establishing their true companionship. 
Our program to carry out NACO K-band spectroscopy of the discovered IR-candidate
companions has been scheduled at the VLT in service mode for the period
October 2005 to March 2006. Using these observations
we will be able to determine the mass of the IR companions much more accurately,
and explore the physics in their atmospheres by comparison of observed and
synthetic spectra. Since the HgMn type primaries have all Hipparcos parallaxes 
($\sigma(\pi)/\pi<0.2$), their age
is known, and assuming coevality we will have an unprecedented
set of data for the confirmed very young or even PMS companions.

We would like to note that our observations contribute not only to the 
understanding of the formation mechanism of HgMn stars but also to the general 
understanding of B-type star formation.
An interesting result about the combination of long- and 
short-period systems has been presented by Tokovinin a few years ago\,\cite{hubrig:to2001}.
He suggested
that the fraction of SBs belonging to multiple systems probably depends on the
SB periods. It is much higher for close binaries with 1 to 10 day periods than
for systems with 10 to 100 day periods. The statistics of multiple systems
is still very poor and much work remains to be done. The proposed survey of
binarity and multiplicity of HgMn stars will help to understand the connection
between close binaries and multiplicity, and especially the formation of close 
binary systems.
To find out which role membership of HgMn stars in multiple systems plays for the 
development of their chemical peculiarities, it would be important to compare the
ranges of periods, luminosity ratios, and orbital eccentricities, as well as 
hierarchy of multiples with the same characteristics of normal late B systems.

A further remarkable feature of HgMn spectroscopic binaries is that many
of them have orbital periods shorter than 20 days. However, binary periods
of less than three days are absent, while they are quite common among 
normal late B systems. Interestingly, from a survey of the Batten et~al.\
catalogue \cite{hubrig:ba1989} limited to systems brighter that V = 7, it appears that
only six normal B8 and B9 stars are known to be members of SBs with orbital 
periods between 3 and 20 days. Four of them are very fast rotating with 
$v\,\sin i$ values of the order of 100\,km\,s$^{-1}$ and more, i.e.\ much faster
than typical HgMn stars.
For the remaining two systems no information on the rotational velocity could be found
in the literature.

In some binary systems with an HgMn primary, the components definitely rotate
subsynchronously\,\cite{hubrig:gu1986}. It is striking that the majority of 
these systems have more than two components. Probably the most intriguing and 
most fundamental question is whether all late-B close binaries with 
subsynchronously rotating companions belong to more complex systems.

%
%

%
%


\printindex
\end{document}